\documentclass[a4paper,10pt,fleqn]{article}
\usepackage[colorlinks,bookmarksopen,bookmarksnumbered,linkcolor=black,citecolor=black,urlcolor=black,breaklinks]{hyperref}
\usepackage{amsmath}
\usepackage{graphicx}
\usepackage{breakurl}
\usepackage[authoryear]{natbib}
\usepackage{times}
\usepackage{mathrsfs}
\usepackage{color}

\newcommand{\dif}{{\,\rm d}}
\newcommand{\erw}{{\,\rm E}}
\newcommand{\eins}{{\bf 1}}

\begin{document}
\title{Design aspects of COVID-19 treatment trials: Improving probability and time of favourable events}

\author{Jan Beyersmann$^1$, Tim Friede$^{2,3}$\footnote{Corresponding author: Tim Friede, Institut f\"ur Medizinische Statistik, Universit\"atsmedizin G\"ottingen, Humboldtallee 32, 37073 G\"ottingen, Germany, {\sf{e-mail: tim.friede@med.uni-goettingen.de}}, Phone: +49-551-39-4991, Fax: +49-551-39-4995}, Claudia Schmoor$^4$ \\ \\
\normalsize $^1$ Institut f\"ur Statistik, Universit\"at Ulm, Ulm, Germany \\
\normalsize $^2$ Institut f\"ur Medizinische Statistik, Universit\"atsmedizin G\"ottingen, G\"ottingen, Germany \\
\normalsize $^3$ Deutsches Zentrum f\"ur Herz-Kreislaufforschung (DZHK), Standort G\"ottingen, G\"ottingen, Germany \\
\normalsize $^4$ Zentrum Klinische Studien, Universit\"atsklinikum Freiburg, Medizinische Fakult\"at, \\  Albert-Ludwigs Universit\"at Freiburg, Freiburg im Breisgau, Germany}

\date{26 November 2020}

\maketitle         

\begin{abstract}
As a reaction to the pandemic of the severe acute respiratory syndrome coronavirus 2 (SARS-CoV-2), a multitude of clinical trials for the treatment of SARS-CoV-2 or the resulting corona disease (COVID-19) are globally at various stages from planning to completion. Although some attempts were made to standardize study designs, this was hindered by the ferocity of the pandemic and the need to set up trials quickly. We take the view that a successful treatment of COVID-19 patients (i) increases the probability of a recovery or improvement within a certain time interval, say 28 days; (ii) aims to expedite favourable events within this time frame; and (iii) does not increase mortality over this time period. On this background we discuss the choice of endpoint and its analysis. Furthermore, we consider consequences of this choice for other design aspects including sample size and power and provide some guidance on the application of adaptive designs in this particular context. 
\end{abstract}

\section{Introduction}
\noindent 
At the time of writing, the severe acute respiratory syndrome coronavirus 2 (SARS-CoV-2) pandemic is ongoing. As a reaction to the pandemic, a multitude of clinical trials on the treatment of SARS-CoV-2 or the resulting corona disease (COVID-19) are globally in planning, were recently initiated or already completed. Although some attempts were made to standardize study designs, this was hindered by the ferocity of the pandemic and the need to set up trials quickly.

For randomized controlled trials evaluating the safety and efficacy of COVID-19 treatments, the discussion on appropriate outcomes has received considerable attention in the meanwhile. For instance, \cite{dodd2020endpoints} discuss the use of survival methodology
  to investigate \emph{both} an increase of the event probability of a
  favorable outcome such as improvement or recovery \emph{and} its
  timing. Similar to \cite{mccaw2020quantify}, Dodd et al.\ stress that such
  outcomes are subject to competing risks or competing events, because
  patients may die without having achieved the favorable outcome. The authors
  argue that these patients must not be censored at their time of death but,
  say, at day~28 after treatment, if the trial investigates a
  28-day-follow-up. The authors continue to advocate investigating hazard
  ratios based on such data (called improvement or recovery rate ratio by the
  authors, because the outcome is not hazardous to health, but favorable). 
	Furthermore, \cite{Benkeser2020}, an early methodological
  publication on COVID-19, consider time-to-event outcomes in a paper
  advocating covariate adjustment in randomized trials, but do not consider
  competing events. Rather, the authors consider a composite of intubation and
  death, thereby avoiding the need to model competing events. This composite
  combines two unfavorable outcomes, but for outcomes recovery or improvement,
  such a composite endpoint that also includes death is not meaningful. 

  The role of censoring here is subtle: \cite{dodd2020endpoints} state that
  the improvement or recovery rate ratio approach coincides with the
  subdistribution hazard ratio approach of \cite{FineGray1999}, \emph{if}
  there is additional, `usual' censoring as a consequence of staggered study
  entry. \cite{mccaw2020quantify}, on the other hand, characterize the
  approach to censor previous deaths at day~28, still assuming a
  28-day-follow-up, as `unusual' and warn against the use of one minus
  Kaplan-Meier, \emph{if} there is additional censoring before day~28, say,
  because of staggered entry. In our presentation of motivating examples in
  Section~\ref{sec:examples}, we will see that both Kaplan-Meier estimates
  censoring at the time of death and Kaplan-Meier estimates censoring at
  day~28 are being used in COVID-19 trials.

 Recently, \citet{Kahan2020} discussed outcomes in the light of the estimand framework. One aim of this paper is to clarify and provide guidance with respect to the estimands at hand when using survival methodology to investigate \emph{both}
  an increase of the event probability of a favorable outcome \emph{and} its
  timing. To this end, we will demonstrate that censoring deaths on day~28 in
  a trial with a 28-day-follow-up conceptually corresponds to formalizing time
  to improvement or recovery via improper failure times, which we will call
  subdistribution times, with probability mass at infinity. The latter
  corresponds to the probability of death during 28-day or, more generally,
  $\tau$-day-follow-up. This has various consequences: It allows to formalize
  mean and median times to improvement (recovery). The Kaplan-Meier estimator
  based on death-censored-on-day-$\tau$-data will coincide with the Aalen-Johansen
  estimator of the cumulative event probability considering the competing event death, provided that there
  is no additional censoring. The hazard ratio at hand will be a
  subdistribution hazard ratio as a consequence of using subdistribution
  times, but not as a consequence of additional censoring.

In this manuscript, we take the view that a successful treatment of COVID-19 patients (i) increases the probability of
a recovery or improvement within a certain time interval, say 28 days; (ii) aims to expedite recovery or improvement within this time frame; and (iii)
does not increase mortality over this time period (see e.g. \citet{Wilt2020}). This should be reflected in the main outcomes of a COVID-19 treatment trial. The choice of outcomes has also some implications for the trial design. Firstly, even in traditional the sample size calculation might be complicated by the presence of competing events. Secondly, novel trial designs including platform trials and adaptive group-sequential designs are more frequently applied than usual in COVID-19 treatment trials. \citet{Stallard2020} provide an overview over such designs, discuss their utility in COVID-19 trials and make some recommendations. In the light of the outcome discussion, we provide some comments on the application of such outcomes in adaptive designs.

The manuscript is organized as follows. In Section \ref{sec:examples}
background on some example trials is provided to motivate the investigations
presented here. In Section \ref{sec:outcomes} outcomes, their analysis and
interpretation are considered before some guidance is provided on planning
such trials in Section \ref{sec:design}. We close with a brief discussion in
Section \ref{sec:discussion}.

\section{Motivating examples} \label{sec:examples}

Our starting point is that a successful treatment of COVID-19
  patients (i) increases the proportion of recoveries within a time interval
  $[0, \tau]$, say, $\tau = 28$ days; (ii) aims to expedite recovery on
  $[0, \tau]$; and (iii) does not increase mortality at time~$\tau$. Aim (i) is
  obviously desirable both from a patient's perspective and from a public
  health perspective. The rationale behind aim (ii) is that two different
  treatments that lead to comparable recovery proportions at time~$\tau$ may
  differ in the timing of recoveries. Here, faster recovery is not only
  desirable from a public health perspective with respect to available
  resources, but faster recovery from ventilation will also benefit the
  individual patient. Finally, requirement (iii) reflects that a treatment that
  increases both the proportion of recoveries and the proportion of deaths at
  time~$\tau$ benefits some patients and harms others.

  We will argue that aims (i)--(iii) cast COVID-19 trials into a competing
  events (or competing risks) setting, although this is not necessarily or not
  explicitly recognized. For example, the primary clinical endpoint of
  \cite{Wang2020} was time to clinical improvement within 28 days after
  randomisation, addressing aims (i) and (ii). Within $\tau = 28$ days, 13\% of
  the patients in the placebo group and and 14\% in the treatment (Remdisivir)
  group died. The authors aimed to address such competing mortality before
  clinical improvement by right-censoring time to clinical improvement
  at~$\tau$ for patients dying before~$\tau$. The authors then used the usual
  machinery of Kaplan-Meier, log-rank and Cox proportional hazards
  regression. However, as we will see below, their analysis amounts to using the Aalen-Johansen
  estimator of the cumulative event probability (instead
  of Kaplan-Meier), Gray's test for comparing cumulative event probabilities
  (or subdistributions) between groups (instead of the
  common log-rank test) and Fine and Gray's proportional subdistribution
  hazards model (instead of the usual Cox model)\citep{bam}.
		
	Other recent examples are \citet{Beigel2020} who consider time
  to recovery on days~$[0, 28]$ and \citet{Cao2020} whose primary outcome is
  time to clinical improvement until day~28. Beigel et al. also censor
  previous deaths `on the last observation day' (see the Appendix of Beigel et
  al.) and use Kaplan-Meier (here, actually, Aalen-Johansen) and log-rank test
  (here, actually, Gray's test) to analyse these data. Cao et al. censor both
  `failure to reach clinical improvement or death before day 28' on day~28 and
  use Kaplan-Meier, log-rank and the Cox model (here, actually, Fine and
  Gray). Interestingly, Cao et al. comment that `right-censoring occurs when
  an event may have occurred after the last time a person was under
  observation, but the specific timing of the event is unknown', although this
  is clearly not the case for patients censored at day~28 following death
  before that time.
  
  To be precise, let~$\vartheta$ be the time to clinical improvement, using
  the example of Wang et al.. Time to improvement is, in general, \emph{not}
  well defined for patients dying prior to improvement. To address this, the
  subdistribution time is defined as $\vartheta=\infty$ for the latter
  patients. The interpretation of the improper random variable~$\vartheta$ is
  that it equals the actual time of improvement when $\vartheta <
  \infty$. However, patients who die before improvement will never experience
  this primary outcome and, hence, $\vartheta=\infty$. The censored
  subdistribution time in the paper of Wang et al.\ becomes
  \begin{displaymath}
    \tilde\vartheta = \min(\vartheta, \tau).
  \end{displaymath}
  Writing~$T$ for the time to improvement ($\varepsilon=1$) or death
  ($\varepsilon=2$), whatever comes first, one minus the Kaplan-Meier
  estimator based on the censored $\tilde\vartheta$ data equals the
  Aalen-Johansen estimator of the cumulative improvement probability
  \begin{displaymath}
    P(T\le t, \varepsilon=1).
  \end{displaymath}
  This has been documented elsewhere \citep{gesk:2010} but it is most easily
  seen for the present case of censoring at one commmon~$\tau$. We provide the
  calculation in the Appendix. In the previous display, $T$ is the
  time-to-first-event and $\varepsilon$ is the type-of-first-event, i.e.,
  $P(T\le t, \varepsilon=1)$ is the cumulative event probability or the
  so-called cumulative incidence function of the 
  type~$1$ event. Censoring at one commmon~$\tau$ is a simple case of `censoring
  complete' data \citep{FineGray1999} which we exploit in the
  Appendix. `Censoring complete' means that $\tilde\vartheta$ is also known
  for those observed to die before improvement. `Observed to die' means that
  death has occurred before censoring. Such patients will not be further
  followed-up, and the future censoring time becomes latent and is, in
  general, not known. This complicates general technical developments in the
  subdistribution framework \citep{FineGray1999}, but here the censored
  subdistribution times are known, see our Appendix. The hazard `attached'
  to~$\vartheta$ neither equals the all-events hazard of~$T$ nor the
  event-specific hazards of~$T$ and either~$\varepsilon=1$ or~$\varepsilon=2$
  \citep{bam}. One important consequence is that using standard log-rank
  software on the censored subdistribution times gives Gray's test for
  comparing event probabilities~$P(T\le t, \varepsilon=1)$ between groups,
  using standard proportional hazards software employs Fine and Gray's
  proportional subdistribution hazards model, casting these analyses into a
  competing events setting.

  We will investigate the consequences of competing events in the following
  sections, including alternatives to the subdistribution framework, the need
  to still analyse competing mortality and possible strategies to account for
  death \emph{after} recovery. Here, we stress that the aim to account for our
  items i)--iii) above has led authors to implicitly employ a competing events
  analysis, although this is not explicitly acknowledged. One worry is that
  the subdistribution hazards framework at hand has repeatedly been
  re-examined, questioning the interpretability of a hazard belonging to an
  improper random variable \citep{andersen2011interpretability}. The key issue
  here is that patients are still kept `at risk' after death and until
  censoring at~$\tau$, although, of course, no further events will be observed
  for these patients.

  There are further examples of the presence of competing events in COVID-19
  studies. For instance, \cite{Grein2020} use Kaplan-Meier for time-to-clinical improvement
  at day~$28$. These authors report a Kaplan-Meier estimate of 84\% for the
  cumulative improvement probability at~$\tau=28$, although improvement was
  only observed for 36 (68\%) out of 53 patients. Letters to the Editor and a
  Reply by Grein et al.\ reveal that the original Kaplan-Meier analysis had
  censored deaths before improvement at the time of death, but not
  at~$\tau$. It is well known that such an analysis is subject to `competing
  risks bias' and must inevitably overestimate cumulative event
  probabilities \citep{bam}.

  As the last example of this section we consider ventilator-free days (VFDs),
  see \cite{Yehya2019}, which is the primary or secondary outcome in a number
  of ongoing trials (trial identifiers NCT04360876, NCT04315948, NCT04348656,
  NCT04357730, NCT03042143, NCT04372628,NCT04389580 at\\
  \texttt{clinicaltrials.gov} and DRKS00021238 at
  \texttt{clinicaltrialsregister.eu}.) Yehya et al.\ provide an applied
  tutorial on using VFDs as an outcome measure in respiratory
  medicine. Similar to our aims (i)--(iii) above, they argue in favour of using
  VFDs, because they `penalize nonsurvivors', that using time as an outcome
  `provide[s] greater statistical power to detect a treatment effect than the
  binary outcome measure' and that time is relevant in that `shortened
  ventilator duration is clinically and economically meaningful'. Again,
  censored subdistribution times are present, although this is not made
  explicit in the definition of VFDs. Choosing once more a time horizon
  of~$\tau=28$ days, VFDs are defined as $28-x$ if ventilation stops on
  day~$x$, but are defined as~$0$ if the patient either dies while being
  ventilated or is still alive and ventilated after~$[0, 28]$. Interpreting
  the subdistribution time~$\vartheta$ as the day when ventilation is stopped
  while alive (and not as a consequence of death), we get
  \begin{displaymath}
    \mbox{VFDs} = 28 - \tilde\vartheta
  \end{displaymath}
  and Yehya et al.\ consequently also suggest the proportional subdistribution
  hazards model as one possible statistical analysis.

\section{Outcomes, their analysis and interpretation} \label{sec:outcomes}

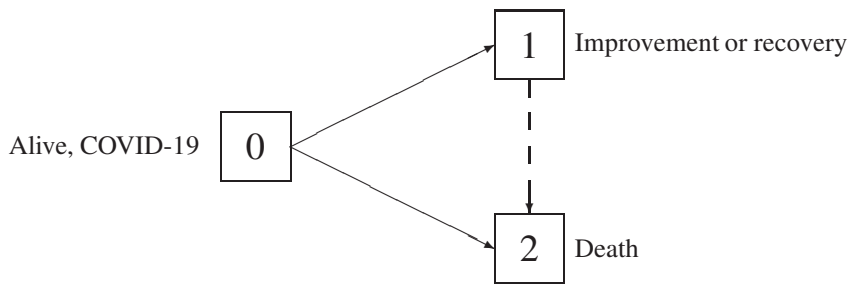
\begin{figure}[htb] \centering
    \setlength{\unitlength}{0.9cm} \begin{picture}(5,5)(0.6,-2)
      \put(0,0){\framebox(1,1){\Large 0}}
      \put(0,0){\makebox(1,1){\parbox{6.5cm}{Alive, COVID-19}}}
      \put(1,0.5){\vector(2,1){3}} \put(4,1.5){\framebox(1,1){\Large 1}}
      \put(1,0.5){\vector(2,-1){3}}\put(4,-1.5){\framebox(1,1){\Large 2}}
      \put(4.5,1.475){\line(0,-1){0.25}}\put(4.5,1){\line(0,-1){0.25}}
      \put(4.5,0.5){\line(0,-1){0.25}}\put(4.5,0){\vector(0,-1){0.5}}
      \put(8,1.5){\makebox(1,1){\parbox{6cm}{Improvement or recovery}}}
      \put(8,-1.5){\makebox(1,1){\parbox{6cm}{Death}}} \end{picture} \caption{Competing
      events model (solid arrows only) and illness-death model without
      recovery (solid and dashed arrows) for outcomes improvement or recovery
      in the presence of the competing event death.}  \label{fig:fig} \end{figure}

Consider a stochastic process~$(X(t))_{t\in[0,\tau]}$ with state
  space~$\{0,1,2\}$, right-continuous sample paths and initial state~$0$,
  $P(X(0)=0)=1$, see Figure~\ref{fig:fig}. Patients alive with COVID-19,
  randomized to treatment arms, are in state~$0$ of the Figure at
  time~$0$.  
	
  Our main model will be the competing events model in Section~\ref{sec:ce},
  considering only the solid arrows in the Figure. Improvement or recovery is
  modelled by a $0\to 1$ transition, death without prior improvement (or
  recovery) is modelled by a $0\to 2$ transition. Later, in
  Section~\ref{sec:idm}, we will briefly consider an extension of this model
  to an illness-death model without recovery by also considering $1\to 2$
  transitions, i.e., death after improvement events. This is illustrated in
  Figure~\ref{fig:fig} by the dashed arrow. Note that `illness-death without
  recovery' does not mean that recovery may not be modelled, but that $1\to 0$
  transitions are not considered. In terms of outcomes, $X(t)=1$ in the
  competing events model means that improvement has occurred on~$[0,t]$, but
  in the illness-death model, $X(t)=1$ means that improvement has occurred
  \emph{and} that the patient is still alive. The distinction may be relevant
  for trials in patients where possible subsequent death on $[0, \tau]$ is a
  concern, see \cite{Sommer2018} for a discussion of death after
  clinical cure in treatment trials for severe infectious diseases.

  \subsection{Competing events: time to and type of first event}\label{sec:ce}
  
  Both in the competing events and, later, in the illness-death model, time-to-first-event is
  \begin{equation}
    \label{eq:T}
    T = \inf\{t\;:\; X(t)\neq 0\},
  \end{equation}
  the waiting time in state~$0$ of the Figure, with type-of-first-event $\epsilon=X(T)$,
  \begin{equation}
    \label{eq:XT}
    X(T)\in\{1,2\},
  \end{equation}
  the state the process enters upon leaving the initial state. The tupel~$(T,
  X(T))$ defines a \emph{competing events} situation. Note that competing
  events are characterized by time-to-first-event and type-of-first-event; it
  is not assumed that there are no further events after a first
  event. However, the analysis of subsequent events requires more complex
  models such as an illness-death model.

  The stochastic process for time and type of the first event is regulated by
  the event- (or cause-) specific hazards
  \begin{equation}
    \label{eq:esh}
    \alpha_{0j}(t) = \lim_{\Delta t \searrow 0}\frac{P(T \in [t, t + \Delta
      t), X(T)=j\,|\, T\ge t)}{\Delta t}, \ j\in\{1,2\}, t\in [0, \tau],
  \end{equation}
  which we assume to exist. Their sum
  \begin{displaymath}
    \alpha_{0\cdot}(t) = \alpha_{01}(t) + \alpha_{02}(t)
  \end{displaymath}
  is the usual all-events hazard of time~$T$ with survival function
  \begin{displaymath}
    P(T>t) = \exp\left(-\int_0^t  \alpha_{0\cdot}(u)\dif u\right).
  \end{displaymath}
  Note that~$T$ is the time until a composite of improvement or death,
  whatever comes first, which is \emph{not} a meaningful outcome in the
  present setting, combining an
  endpoint that benefits the patient with one that harms the patient. Rather,
  as discussed above, authors consider the cumulative improvement probability
  \begin{eqnarray}
    \label{eq:cif}
    F_1(t) = P(T \le t, X(T) = 1) & = & \int_0^t P(T\ge u) \alpha_{01}(u) \dif u\\
    {} & = & 1 - P(\vartheta > t),
  \end{eqnarray}
  with subdistribution time~$\vartheta$ until improvement defined as
  \begin{displaymath}
    \vartheta = \left\{
      \begin{array}{r@{\quad \mbox{if} \quad}l}
        T & X(T) = 1\\
        \infty & X(T) = 2.
      \end{array}
    \right.
  \end{displaymath}
  A binary outcome, say improvement status at time~$\tau$, is covered by this framework,
  \begin{displaymath}
    \eins(T\le \tau, X(T)=1) = \left\{
      \begin{array}{r}
        1\\
        0 
      \end{array}
    \right.
  \end{displaymath}
  with indicator function~$\eins(\cdot)$ and improvement probability (at
  time~$\tau$) $P(T \le \tau, X(T) = 1)$. However, viewing
  quantities~\eqref{eq:cif} as a function of time allows to detect earlier
  improvement with possibly comparable improvement probabilities
  at~$\tau$. The expected proportion of deaths at~$\tau$ without prior
  improvement is $P(T \le \tau, X(T) = 2)$.

  Key quantities of the competing events model are time and type of first
  event, $(T, X(T))$, and the event-specific hazards
  $(\alpha_{01}(t), \alpha_{02}(t))$. The subdistribution time~$\vartheta$
  appears to be little more than an afterthought of~\eqref{eq:cif}. However,
  its relevance is closely connected to the subdistribution \emph{hazard}
  which neither equals any of the event-specific hazards nor the all-events
  hazard. It also reappears in the context of `average'
    improvement or recovery times discussed towards the end of the present
    Subsection.

  The subdistribution hazard~$\lambda(t)$ is the hazard `attached'
  to~\eqref{eq:cif} by requiring
  \begin{equation}
    \label{eq:sh}
    P(T \le t, X(T) = 1) = 1 - \exp\left(-\int_0^t \lambda(u)\dif u\right),
  \end{equation}
  leading \citep{bam} to
  \begin{equation}
    \label{eq:jbms}
    \lambda(t) = \left(1 + \frac{P(T \le t, X(T) = 2)}{P(T>t)}\right)^{-1}\alpha_{01}(t),
  \end{equation}
  which illustrates why interpretation of the subdistribution hazard as a
  hazard has been subject of debate \citep{andersen2011interpretability}. The
  event-specific hazards~$\alpha_{0j}(t)$ have the interpretation of an
  instantaneous `risk' of a type~$j$ event at time~$t$ given one still is in
  state~$0$ just prior to time~$t$. They may be visualized
  \citep{beyersmann2014incidence} as forces moving along the solid arrows in
  Figure~\ref{fig:fig}. There is no such interpretation for the
  subdistribution hazard. The above display also illustrates that a competing
  subdistribution hazard, `attached' to $P(T \le t, X(T) = 2)$ may not be
  chosen or modelled freely. This is in contrast to the event-specific
  hazards.

  The rationale of the subdistribution hazard is that it reestablishes a
  one-to-one correspondence with the cumulative event
  probability~\eqref{eq:cif}, which is a function of both event-specific
  hazards~$\alpha_{01}(t)$ and~$\alpha_{02}(t)$. The subdistribution hazard
  approach may also be viewed via a transformation model for~\eqref{eq:cif}
  using the link $x \mapsto \log(-\log(1-x))$ as in a Cox model for the
  all-events hazard, but still without a common hazard interpretation. Several
  authors have argued in favor of link functions which are more amenable to
  interpretation such as a logistic link. See also Section \ref{sec:power} on study planning with respect to odds ratios. Against the background of the motivating examples in
  Section~\ref{sec:examples}, we will put a certain emphasis on the former
  link but also note that it is not uncommon that results from either link
  function coincide from a practical point of view
  \citep{beyersmann13:_compet}. In the absence of competing events, this has
  been well documented for studies with short follow-up and low cumulative
  event probability \citet{Annesi1989}. In the present setting, trials will aim at increasing
  cumulative improvement or recovery probabilities, but `competing' mortality
  implies that these probabilities must be below one, which distinguishes
  competing events from the all-events framework.

  Finally, the subdistribution time~$\vartheta$ is useful for
    formalizing `average' improvement or recovery times. Assuming `competing'
    mortality, i.e., $P(T \le \tau, X(T) = 2) > 0$, it is easy to see that
    \begin{displaymath}
      \erw(\vartheta) = \infty,
    \end{displaymath}
    because~$P(\vartheta = \infty) = P(X(T) = 2)$. Consequently, the expected
    or mean time to improvement (recovery) is not a useful parameter. It is
    well known that standard survival analysis (time-to-all-causes-death), expected
    survival time is typically not investigated, but for a different reason. For
    the latter, expected survival time is a finite number, but it is usually not
    identifiable, at least not nonparametrically, because of limited
    follow-up. Both in this and in the present context, two possible solutions
    for investigating `average' times-to-event are restricted means and median
    times. For the former, \cite{andersen2013decomposition} considers
    \begin{equation}
      \label{eq:lyl}
      \erw\left(\min(\vartheta, \tau)\right) = \tau - \int_0^\tau F_1(u)\dif u.
    \end{equation}
    If the competing events are different causes of death, Andersen
    interprets~\eqref{eq:lyl} as the mean time span lost before time~$\tau$
    and `due to cause~1'. The area under the cumulative event probability
    \begin{displaymath}
      [0, \tau] \ni t \mapsto F_1(t)
    \end{displaymath}
    may hence be interpreted as the mean time to improvement (recovery) before
    time~$\tau$. For COVID-19 trials, this parameter has recently been
    suggested by \cite{mccaw2020selecting}, however, without giving formulae
    or making the link to subdistribution times explicit. This is also related to the ventilator free days discussed in Section \ref{sec:examples}. A recent example of
    using~\eqref{eq:lyl} is \cite{HAO2020354} who considered
    influenza-attributable life years lost before the age of~90.  

    Alternatively, one may consider the median time to improvement,
    \begin{equation}
      \label{eq:median}
      \inf\left\{t\, :\, F_1(t)\ge 0.5\right\}.
    \end{equation}
    Again, there is a conceptual difference to median survival time in that the
    latter always is a finite number (denying the possibility of immortality),
    but quantity~\eqref{eq:median} will be defined as infinity, if the
    eventual improvement probability does not reach~50\%. However, if its
    Aalen-Johansen estimator (see Subsection~\ref{sub:stat} below) does
    reach~50\% on~$[0,\tau]$, the median time-to-improvement can be estimated
    nonparametrically by plug-in of the Aalen-Johansen estimator
    in~\eqref{eq:median}, see \cite{Beye:Schu:note} for technical details. Use
    of~\eqref{eq:median} as an end point in COVID-19 treatment trials
    accounting for competing events has recently been considered
    by~\cite{mccaw2020quantify}. Study planning of some recent randomized clinical trials on treatment of COVID-19 was also based on assumptions on median times to clinical improvement (\cite{Cao2020}, \cite{Li2020}). This will be discussed in Section~\ref{sec:power}. We will use improvement and recovery interchangeably from a methodological perspective as examples for favourable events (although clinically they are of course different).

  \subsection{Statistical approaches}\label{sub:stat}
  We will assume that follow-up data are complete in that improvement status
  and vital status are known for all patients on~$[0, \tau]$. In most of the
  motivating examples of Section~\ref{sec:examples}, both patients alive but
  without improvement up to day~$\tau$ and patients who had died were censored
  at~$\tau$. Although censored at this maximal time point, both improvement
  status and vital status are known for these patients on~$[0,
  \tau]$. Survival methodology discussed below will allow for right-censoring
  of patients alive where further follow-up information ceases at the time of
  censoring, but we assume that this is a minor problem when, e.g.,
  $\tau = 28$ days.

  Assuming data to be complete in this sense, the Kaplan-Meier estimator
  of~$P(T>t)$ is
  \begin{displaymath}
    \hat P(T>t) = \prod_{u \le t}\left(1-\frac{\Delta N(u)}{Y(u)}\right) = 1 -
    \frac{N(t)}{n},
  \end{displaymath}
  where the product in the above display is over all unique event times,
  $N(t)$ is the number of composite events (transitions out of the initial
  state in Figure\ref{fig:fig}) on~$(0,t]$, $\Delta N(t)$ is the number of
  events at time~$t$, $Y(t)$ is the number at risk just prior time~$t$ and $n$
  is the sample size. Because of complete follow-up on $[0, \tau]$,
  $\hat P(T>t)$ equals the empirical event-free fraction $(n - N(t))/n$.

  Introducing
  \begin{eqnarray*}
    N_{01}(t) & = & \mbox{no. of~$0\to 1$ transitions on~$(0,t]$},\\
    N_{02}(t) & = & \mbox{no. of~$0\to 2$ transitions on~$(0,t]$},\\
    N(t) & = & N_{01}(t) + N_{02}(t),
  \end{eqnarray*}
  with $\Delta N_{0j}(t)$ type~$j$ events precisely at time~$t$, the
  Aalen-Johansen estimators are
  \begin{displaymath}
    \hat P(T\le t, X(T)=j) = \sum_{u \le t} \hat P(T\ge u) \frac{\Delta
      N_{0j}(t)}{Y(t)} = \frac{N_{0j}(t)}{n},\ j\in\{1,2\},
  \end{displaymath}
  which also equal the usual empirical proportions assuming complete
  follow-up. Here, one easily sees that
  \begin{displaymath}
    \hat P(T>t) + \hat P(T\le t, X(T)=1) + \hat P(T\le t, X(T)=2) = 1,
  \end{displaymath}
  a natural balance equation which is maintained even in the presence of
  censoring, but violated if one were to use
  \begin{equation}
    \label{eq:KMboese}
    1 - \prod_{u \le t}\left(1-\frac{\Delta N_{01}(u)}{Y(u)}\right)    
  \end{equation}
  to estimate the cumulative probability of a type~$1$ event, see the
  motivating examples of Section~\ref{sec:examples}. This Kaplan-Meier-type
  estimator inevitably overestimates, the reason being that one minus
  Kaplan-Meier approximates an empirical distribution function, but the
  cumulative probability of a type~$1$ event is bounded from above
  by~$P(X(T)=1)$.

  However, the Kaplan-Meier estimator predominantly used in the motivating
  examples of Section~\ref{sec:examples} is based on subdistribution times
  with censoring time~$\tau$ leading to
  \begin{eqnarray*}
    \tilde N_{01}(t) & = & \sum_i^n \eins\left(\vartheta_i \le t\right), \\
    \tilde Y(t) & = & Y(t) + \sum_i^n \eins\left(T_i < t, X(T_i) = 2\right), t \le \tau,
  \end{eqnarray*}
  where index~$i$ signals patient~$i$, $i\in\{1, \ldots n\}$. Because
  \begin{displaymath}
    \eins\left(\vartheta_i \le t\right) = \eins\left(T_i \le t, X(T_i)=1\right)
  \end{displaymath}
  we have $\tilde N_{01}(t) = N_{01}(t)$, but $\tilde Y(t) \ge Y(t)$, because
  censoring dead patients at~$\tau$ enlarges the risk set by the number of
  previous deaths. In the current setting, it is easy to demonstrate that
  \begin{equation}
    \label{eq:yeah}
    \hat P(T\le t, X(T)=j) = 1 - \prod_{u \le t}\left(1-\frac{\Delta
        N_{01}(u)}{\tilde Y(u)}\right),
  \end{equation}
  see the Appendix. Note that the difference between the right hand side of
  \eqref{eq:yeah} and the biased Kaplan-Meier-type
  estimator~\eqref{eq:KMboese} lies in the use of a different risk set.

  Any regression model for hazards may be fit to the event-specific hazards,
  the most common choice being Cox models,
  \begin{displaymath}
    \alpha_{0j}(t; Z) = \alpha_{0j;0}(t)\cdot \exp\left(\beta_{0j}^\top
      Z\right),\ j\in\{1,2\},
  \end{displaymath}
  with event-specific baseline hazards~$\alpha_{0j;0}(t)$, event-specific
  $p\times 1$ vectors of regression coefficients~$\beta_{0j}$ and a
  $p\times 1$ vector of baseline covariates~$Z$. Technically, an
  event-specific Cox model for the type~$1$ hazard, say, may be fit by only
  counting type~$1$ events as events and by additionally censoring type~$2$
  events at the time of the type~$2$ event. Roles reverse fitting an
  event-specific Cox model for the type~$2$ hazard. Interpretationally, this
  has arguably been a source of confusion, because the biased
  Kaplan-Meier-type estimator~\eqref{eq:KMboese} also only counts type~$1$
  events as events and additionally censors type~$2$ events. The difference
  between fitting an event-specific Cox model and the Kaplan-Meier-type
  estimator~\eqref{eq:KMboese} is that \emph{hazard} models allow for quite
  general censoring processes including censoring by a competing
  event. However, \emph{probabilities} depend on all event-specific hazards,
  which is why we have formulated Cox models for all event types
  above. It is, however, not uncommon to only see results from
    one event-specific Cox model being reported,
    see~\cite{Goldman2020,spinner2020effect} for two recent examples from
    COVID-19 treatment trials.

    In contrast to, e.g., these two studies, event-specific Cox models have
  not been used in the motivating examples above. Rather a Cox-type model, the
  Fine and Gray model, for the subdistribution hazard has been employed,
  \begin{displaymath}
    \lambda(t; Z) = \lambda_{0}(t)\cdot \exp\left(\gamma^\top Z\right).
  \end{displaymath}
  If the cumulative improvement probabilities follow the Fine and Gray model,
  a subdistribution hazard ratio larger than one for treatment signals both an
  increase of the expected improvement proportion at~$\tau$ and earlier
  improvement.

  It has been repeatedly argued that any competing events analysis should
  consider all competing events at hand. For the event-specific hazards, we
  have therefore formulated two Cox models. For the Fine and Gray approach,
  postulating a Cox-type model for the `competing' subdistribution hazard is
  complicated by~\eqref{eq:jbms}. However, delayed death on~$[0,\tau]$ does
  not benefit the patient if~$\tau=28$ days. Hence, in the present setting, it
  will suffice to consider the probability~$P(X(T)=2)$ of `competing'
  probability by common methods for proportions. 

  \subsection{Death after improvement or recovery: illness-death
    model}\label{sec:idm}
  For instance, \cite{mccaw2020quantify} broach the issue of
    longer follow-up in future COVID-19 treatment trials and its impact on
    meaningful outcomes including time-to-death. Here, one aspect is that
    prolonged survival on~$[0, \tau]$, where, e.g., $\tau$ is 28 days, does not
    benefit patients \citep{Tan2020}. The aim of the present subsection is to briefly
    outline how the competing events framework may be extended to also handle
    death events possibly after improvement or recovery during a longer
    follow-up. To this end, define for finite times~$t$ the transition hazard
    \begin{equation}
      \label{eq:idm}
      \alpha_{12}(t;\vartheta) = \lim_{\Delta t \searrow 0}\frac{P(X(t + \Delta
        t)=2\,|\, X(t-)=1, \vartheta<t)}{\Delta t},
    \end{equation}
    where we now also model $1\to 2$ transitions along the dashed arrow in
    Figure~\ref{fig:fig}. The model has recently been used to jointly model
    time-to-progression (not a favorable outcome, of course) or
    progression-free-survival and overall survival by \cite{meller2019}. The
    model is time-inhomogeneous Markov, if $\alpha_{12}(t;\vartheta)$ does not
    depend on the finite value of~$\vartheta$. Again a proportional hazards
    model may be fit to the transition hazard, possibly also modelling
    departures from the Markov assumption, but the interpretation of
    probabilities arguably is more accesible. One possible outcome could be
    the probability to be alive after recovery, i.e., $P(X(t)=1)$ over
    relevant time regions. In the context of clinical trials such outcomes
    have recently been advocated by \cite{Sommer2018} for treatment
    trials for severe infectious diseases and by \cite{bluhmki2020relapse} for
    patients after stem cell transplantation whose health statuses may switch
    between favorable and less favorable. \cite{schmidt2020extracorporeal} have recently used such a multistate
    model in a retrospective cohort study on COVID-19 patients, modelling
    oxygenation and intensive care statuses. For the statistical analysis, the
    authors used both Cox models of the transition hazards and reported
    estimated state occupation probabilities and `average' occupation times.

\section{Some design considerations} \label{sec:design}
Following on from the consideration of the choice of outcomes, their analysis and interpretation in COVID-19 trials, we now look into the consequences a particular choice of outcome has for the design of the trial. We start with sample size considerations and then comment on the use of adaptive designs.

\subsection{Power and sample size considerations} \label{sec:power}
For a randomized clinical trial for the investigation of the effect of a COVID-19 treatment on clinical improvement or recovery of patients or death, various approaches are conceivable. As described above, the time horizon considered is usually short, often 28 days. So, it can be assumed that the recording of the interesting outcomes as hospitalization, ventilation, clinical symptoms, and death is complete. In this situation, an ordered categorical endpoint as the eight-point ordinal scale proposed in the master protocol of the WHO \citep{who2020} and e.g. used in a seven-point version in the trial by \citet{Goldman2020} at a pre-specified time point (e.g. 28 days) or a simpler binary endpoint as e.g. death or clinical recovery as defined by a dichotomized version of the ordinal scale can be used, as e.g. done by \citet{Lee2020}. An ordered categorical endpoint might be analyzed, under the proportional odds assumption, with a proportional odds model, for which sample size planning can be based on the formula proposed by \citet{Whitehead1993}. Under more general assumptions, the treatment groups might be compared with respect to an ordinal outcome by a nonparametric rank-based approach using e.g. the Wilcoxon rank sum test and the so-called probabilistic index or relative effect \citep{Kieser2013}. The sample size can then be calculated using the formula provided by \citet{Noether1987} or subsequent refinements using the variance under the alternative \citep{Vollandt1997} or extensions to a variety of alternative hypotheses \citep{Happ2019}. A binary endpoint would usually be analyzed with a logistic regression model for which sample size planning can be based on formula (2) in \citet{Hsieh1998}. 

Even if the recording of the interesting endpoint can be assumed to be complete, it may be desirable to analyze not just the occurrence of the endpoint within the specified time period, but the time to the occurrence of the endpoint for mainly two reasons. First, as described in Section \ref{sec:examples}, a time-to-event analysis captures not only a difference between treatments with respect to the proportion of patients for whom the event had occurred, but also a difference between treatments with respect to the time of occurrence. This can be relevant even on a short time interval when the endpoint is e.g. time under mechanical ventilation, which has adverse effects on patients’ health the longer it is required. Second, even if completeness of data over the interesting time period is assumed, individual patients might be lossed to follow-up, which can be handled by a time-to-event analysis being able to include censored observations.

In time-to-event analyses we can model the effect of a treatment on the
(event-specific) hazard (\ref{eq:esh}) or on the cumulative event
probability (\ref{eq:cif}) of experiencing the event. For the planning of
clinical trials with time-to-event endpoints one has to distinguish if the
event of interest is all-encompassing in the sense that every patient will
experience it at some point in time (although potentially after study end), as
e.g. all-cause mortality, or if the observation of the interesting event, as
e.g. improvement or recovery may be precluded by competing events, as
e.g. death without prior recovery.

In the first case, we do not need to make a decision, whether we are mainly interested in the effect of treatment on the hazard or on the cumulative event probability, as comparisons with respect to hazards, usually performed by the logrank test or the Cox proportional hazards regression model, and comparisons with respect to cumulative event probabilities, usually estimated by the Kaplan-Meier method, are equivalent in this situation.

In the second case of an interesting event for which competing events exist, however, the one-to-one correspondence between hazard and cumulative event probability no longer holds. In this situation, not one hazard but two hazards, the so-called event-specific hazards, one for the interesting event and one for the competing event, exist. As described in Section \ref{sec:ce} the cumulative event probability (\ref{eq:cif}) depends on both event-specific hazards. An analysis of the treatment effect on the event-specific hazards consists of two analyses, one for each event-specific hazard. The analyses of the event-specific hazards can be performed with Cox proportional event-specific hazards regression models where in each analysis the time to the interesting event is censored at the time when the competing event occurs. For the analysis of the treatment effect on the cumulative event probability, the most popular method is the Fine and Gray model \citep{FineGray1999} which is a proportional hazards model for the subdistribution hazard which is the hazard ‘attached’ to the cumulative event probability as described in Section \ref{sec:ce}. 

From the fact, that the cumulative event probability depends on both event
specific hazards the following conclusions can immediately be drawn
\citep{bam}. If treatment as compared to control leads to a decrease (or
increase) in the cumulative probability of the interesting event, this can
have two reasons. It can be due to a direct (e.g. physiological) effect of
treatment on the event-specific hazard of the interesting event
or it can be due to an increase (or decrease) the treatment exhibits on the
event-specific hazard of the competing event. Based on the
analysis of the cumulative probability of the interesting event alone, it is
difficult to understand the treatment mechanism leading to a difference in
event probabilities between treatment and control groups, since various
treatment mechanisms can lead to the same difference in event
probabilities. As a consequence, it is usually recommended to conduct three
analyses for a complete understanding of treatment mechanisms, namely
comparisons between treatment and control with respect to the event-specific
hazard of the interesting event, the event-specific hazard of the competing
event, and the cumulative event probability of the interesting event
\citep{Latouche2013}.

In the planning of a clinical trial, one usually has to pre-specify one treatment effect to be analyzed by one primary analysis \citep{Baayen2019}. In the following we discuss the different approaches of focusing on the event-specific hazard or on the cumulative event probability for the situation of our competing events model in Figure \ref{fig:fig} where the interesting event is recovery from COVID-19 and the competing event is death without prior recovery.

For a comparison of treatment groups with respect to the event-specific hazards, the parameter of interest is the event-specific hazard ratio 
\begin{displaymath}
\theta_{ES}=\alpha_{01T}(t)/\alpha_{01C}(t)
\end{displaymath}
with $\alpha_{01T}(t)$ denoting the event-specific hazard of the treatment
group, and
$\alpha_{01C}(t)$ denoting the event-specific hazard of the control group. For
a comparison of treatment groups with respect to the cumulative probability of
the interesting event, the parameter of interest is the subdistribution hazard
ratio
\begin{displaymath}
\theta_{SD}=\log(1-F_{1T}(t))/\log(1-F_{1C}(t)),
\end{displaymath}
which follows from (\ref{eq:sh}) under the assumption of proportional subdistribution hazard functions, with $F_{1T}(t)$ denoting the cumulative probability of the interesting event in the treatment group, and $F_{1C}(t)$ denoting the cumulative probability of the interesting event in the control group.

In our situation, where the event of interest is recovery, i.e. something favourable, for both quantities $\theta_{ES}$ and $\theta_{SD}$ superiority of treatment versus control is represented by a value larger than 1.

Whatever the planned analysis, i.e. analysis of the event-specific hazard
ratio $\theta_{ES}$ or analysis of the subdistribution hazard ratio
$\theta_{SD}$, sample size planning for a two-sided level $\alpha$ test with
power $1-\beta$ under an assumed hazard ratio $\theta$ is
  typically based on the Schoenfeld formula
\citep{schoenfeld1981asymptotic,Latouche2004,ohneberg2013sample,Tai2015} for the total number of required recovery events
\begin{equation}\label{eq:E}
E = (u_{1-\alpha/2}+u_{1-\beta})^2 / [p (1-p) (\log \theta)^2]
\end{equation}
with $p$ denoting the probability of being in treatment group T, and $u_{1-\gamma}$ denoting the $(1-\gamma)$-quantile of the standard normal distribution. The total number of patients to be randomized can then be calculated as $N = E / \Psi$, where $\Psi$ denotes the probability of observing a recovery event. In the absence of censoring, as assumed in our situation of a short planned trial duration of let say 28 days, $\Psi$ can be calculated as
\begin{equation}\label{eq:Psi}
\Psi = p F_{1T}(28) + (1-p) F_{1C}(28) .
\end{equation}
Although for the analysis of the event-specific hazard ratio and the analysis
of the subdistribution hazard ratio the same formula for sample size
calculation is often used, sample size planning, statistical
analyses, and interpretation of results are different, as $\theta_{ES}$ and
$\theta_{SD}$ represent different parameters as described
above. Another issue is that Schoenfeld's formula assumes
  identical censoring distributions in the treatment groups, see
  \cite{schoenfeld1981asymptotic}. This assumption is well justified for time
  to an all-encompassing endpoint and, technically, it lends itself to a
  particularly simple approximation of the covariation process of the logrank
  statistic underlying Schoenfeld's formula. It does, however, have further
  implications in the presence of competing events.

We will illustrate this for the simplistic assumption of constant
event-specific hazards of experiencing the interesting event
recovery in treatment and control groups, $\alpha_{01T}$ and $\alpha_{01C}$,
and of experiencing the competing event death without prior recovery in
treatment and control groups, $\alpha_{02T}$ and $\alpha_{02C}$. Hence, the
event-specific hazard ratios of recovery and of death without prior recovery
are then given by $\theta_{ES} = \alpha_{01T} / \alpha_{01C}$ and
$\theta_{ES-CE} = \alpha_{02T} / \alpha_{02C}$. Under the constant hazards
assumption, the cumulative probability of recovery in treatment group $k$,
$k=T,C$, at time $t$ is given by
\begin{equation} \label{eq:cp_recover}
  F_{1k}(t) = \frac{\alpha_{01k}}{\alpha_{01k} + \alpha_{02k}}  \left[1 - \exp\left(-\left(\alpha_{01k} + \alpha_{02k}\right) t\right)\right] .
\end{equation}
and the cumulative probability of death without prior recovery in treatment group k, k=T,C, at time t is given by 
\begin{equation} \label{eq:cp_death}
    F_{2k}(t) = \frac{\alpha_{02k}}{\alpha_{01k} + \alpha_{02k}}  \left[1 - \exp\left(-\left(\alpha_{01k} + \alpha_{02k}\right) t\right)\right] .
\end{equation}

Table \ref{tab:n} shows for different scenarios of assumed
event-specific hazards of recovery and death in treatment and
control groups and associated event-specific hazard ratios of recovery and
death, the resulting cumulative event probabilities at time point 28 days and
the resulting subdistribution hazard ratios at time point 28
  days. Parameters were chosen to reflect similar scenarios as present in the
recently published randomized clinical trials on COVID-19 therapies, where
observed probabilities of recovery were around 0.5 to 0.8 and observed
probabilities of mortality were around 0.15 to 0.25 \citep{Beigel2020,
  Cao2020, Li2020, Wang2020}.

\begin{table}[ht]
  \caption{Event-specific hazard ratios and the
    subdistribution hazard ratio at time 28 with respect to recovery for different
    scenarios under the constant hazard assumption} \label{tab:hr}
\begin{tabular}{rrrrrrrrrrr}
\hline \hline
$\alpha_{01T}$ & $\alpha_{01C}$ & $\alpha_{02T}$ & $\alpha_{02C}$ &	$\theta_{ES}$ & $\theta_{ES-CE}$ & $F_{1T}(28)$ & $F_{1C}(28)$ & $F_{2T}(28)$ & $F_{2C}(28)$ & $\theta_{SD}(28)$ \\
\hline
0.04 &	0.04 &	0.01 &	0.01 &	1.00 &	1.00 &	0.60 &	0.60 &	0.15 &	0.15 &	1.00 \\
0.04 &	0.04 &	0.01 &	0.02 &	1.00 &	0.50 &	0.60 &	0.54 &	0.15 &	0.27 &	1.18 \\
0.04 &	0.04 &	0.02 &	0.01 &	1.00 &	2.00 &	0.54 &	0.60 &	0.27 &	0.15 &	0.85 \\
0.06 &	0.04 &	0.01 &	0.01 &	1.50 &	1.00 &	0.74 &	0.60 &	0.12 &	0.15 &	1.44 \\
0.06 &	0.04 &	0.01 &	0.02 &	1.50 &	0.50 &	0.74 &	0.54 &	0.12 &	0.27 &	1.71 \\
0.06 &	0.04 &	0.02 &	0.01 &	1.50 &	2.00 &	0.67 &	0.60 &	0.22 &	0.15 &	1.20 \\
0.08 &	0.04 &	0.01 &	0.01 &	2.00 &	1.00 &	0.82 &	0.60 &	0.10 &	0.15 &	1.84 \\
0.08 &	0.04 &	0.01 &	0.02 &	2.00 &	0.50 &	0.82 &	0.54 &	0.10 &	0.27 &	2.17 \\
0.08 &	0.04 &	0.02 &	0.01 &	2.00 &	2.00 &	0.75 &	0.60 &	0.19 &	0.15 &	1.51 \\
0.04 &	0.06 &	0.01 &	0.01 &	0.67 &	1.00 &	0.60 &	0.74 &	0.15 &	0.12 &	0.69 \\
0.04 &	0.06 &	0.01 &	0.02 &	0.67 &	0.50 &	0.60 &	0.67 &	0.15 &	0.22 &	0.83 \\
0.04 &	0.06 &	0.02 &	0.01 &	0.67 &	2.00 &	0.54 &	0.74 &	0.27 &	0.12 &	0.59 \\
0.04 &	0.08 &	0.01 &	0.01 &	0.50 &	1.00 &	0.60 &	0.82 &	0.15 &	0.10 &	0.54 \\
0.04 &	0.08 &	0.01 &	0.02 &	0.50 &	0.50 &	0.60 &	0.75 &	0.15 &	0.19 &	0.66 \\
0.04 &	0.08 &	0.02 &	0.01 &	0.50 &	2.00 &	0.54 &	0.82 &	0.27 &	0.10 &	0.46 \\
\hline \hline
\end{tabular}
\end{table}

Note that the aim of the Table is to illustrate possible
  constellations of the situation at hand, including some for which one would
  not plan a trial. To illustrate, when the event-specific
recovery hazards in treatment and control are identical $(\theta_{ES} = 1)$, a
decreasing effect of treatment as compared to control on the
event-specific death hazard $(\theta_{ES-CE} < 1)$ leads to an
increased cumulative recovery probability
$(\theta_{SD}(28) > 1)$, whereas an increasing effect of
treatment as compared to control on the event-specific death
hazard $(\theta_{ES-CE} > 1)$ leads to a decreased cumulative recovery
probability $\theta_{SD}(28) < 1$. Clearly, one
  would not plan a trial assuming the latter scenario, but it does illustrate
  that any competing events analysis is incomplete without a look at the
  competing event.

It is tempting to compare the magnitudes of $\theta_{ES}$ and
  $\theta_{ES-CE}$ with that of $\theta_{SD}(28)$. A situation of particular
  interest not just for this comparison arises when there is no treatment
  effect on the competing event-specific hazard ratio, $\theta_{ES-CE}=1$. To
  begin, recall that any event-specific hazards analysis is performed by
  handling observed competing events of the other type as censorings. Hence,
  assuming $\theta_{ES-CE}=1$ complies with the assumption of equal censorings
  mechanisms in the groups for using Schoenfeld's formula. Next, a
  proportional subdistribution hazards model will, in general, be misspecified
  assuming proportional event-specific hazards as a consequence
  of~\eqref{eq:jbms}. However, it has been repeatedly noted that
  $\hat \theta_{ES} \approx \hat \theta_{SD}$ if
  $\hat \theta_{ES-CE}\approx 1$
  \citep{j06b:_iscb07,saadati2017prediction}. This is mirrored in the Table,
  in that scenarios with $\theta_{ES-CE}=1$ find comparable values
  of~$\theta_{ES}$ and~$\theta_{SD}(28)$. Note, however,
  that~$\hat \theta_{SD}$ will estimate a time-averaged subdistribution hazard
  ratio, averaged over the whole time span, computation of which requires
  numerical approximations \citep{beyersmann08:_simul}.

  Equality~\eqref{eq:jbms} also illustrates that event-specific and
  subdistribution hazards operate on different scales, and many authors have
  argued that the subdistribution hazard scale is more difficult to
  interpret, see \cite{andersen2011interpretability} for an in-depth
  discussion. We therefore refrain from further comparing the magnitudes of
  the different effect measures and rather continue with considering their
  impact on sample sizes following from Schoenfeld's formula.

For sample size planning of clinical trials where competing events exist,
assumptions are usually based on the expected cumulative event probabilities
\citep{Schulgen2005, Latouche2013, Baayen2019, Tai2015}. Under the constant
event-specific hazards assumption for both the recovery as well
as the death without prior recovery hazard, the underlying hazards can be
calculated from the cumulative event probabilities via equations
(\ref{eq:cp_recover}) and (\ref{eq:cp_death}) as proposed by
\citet{Pintilie2002, Schulgen2005, Baayen2019, Tai2015}.

Table \ref{tab:n} contrasts for some scenarios of cumulative event
probabilities similar to those of some recently published randomized clinical
trials on COVID-19 therapies the corresponding subdistribution recovery hazard ratio
versus the event-specific recovery hazard ratio calculated from
the cumulative event probabilities under the constant
event-specific hazards assumption. Additionally it is shown,
which sample sizes would result if planning addresses the subdistributon recovery
hazard ratio, the event-specific recovery hazard ratio, or the
odds ratio (of the binary endpoint recovery until day 28) for a randomized
clinical trial which aims to show superiority of treatment as compared to
control with respect to recovery from COVID-19 with two-sided type I error of
0.05 and power 0.8.

\begin{table}[ht]
  \caption{Subdistribution recovery hazard ratio and odds ratio
    (OR) at time~28, and event-specific hazard ratio derived from
    cumulative event probabilities under the constant
    event-specific hazard assumption and resulting sample size
    when chosen as parameter for study planning with two-sided type I error
    rate of 0.05 and power 0.8. } \label{tab:n}
\begin{tabular}{rrrrrrrrrrr}
  \hline \hline
  $F_{1T}(28)$ & $F_{1C}(28)$ & $F_{2T}(28)$ & $F_{2C}(28)$ & $\theta_{ES}$	& $N_{ES}$ & $\theta_{ES-CE}$ & $\theta_{SD}(28)$ & $N_{SD}$ & OR (28) & $N_{OR}$ \\
  0.7 &	0.55 &	0.10 &	0.10 &	1.59 &	237 &	1.25 &	1.51 &	300 &	1.91 &	325 \\
  0.7 &	0.55 &	0.15 &	0.15 &	1.65 &	200 &	1.30 &	1.51 &	300 &	1.91 &	325 \\
  0.7 &	0.55 &	0.20 &	0.20 & 	1.76 &	157 &	1.38 &	1.51 &	300 &	1.91 &	325 \\
  0.7 &	0.55 &	0.10 &	0.20 &	1.39 &	474 &	0.54 &	1.51 &	300 &	1.91 &	325 \\
  0.7 &	0.55 &	0.15 &	0.20 &	1.54 &	274 &	0.91 &	1.51 &	300 &	1.91 &	325 \\ 
  \hline \hline
\end{tabular}
\end{table}

Table~\ref{tab:n} invites some discussion. To begin, we reiterate that
  Schoenfeld's formula assumes identical censoring mechanisms in the treatment
  groups. This is formally fullfilled when planning an analysis
  of~$\theta_{ES}$ when $\theta_{ES-CE}=1$. In this case, a beneficial
  (harmful) effect on~$\theta_{ES}$ directly translates into a beneficial
  (harmful) effect on the cumulative recovery probability. If the assumption
  of identical censoring mechanisms is violated, the reported sample sizes
  should serve as a starting point for simulation based sample size planning
  in practice. For the subdistribution approach, \citet{Latouche2004}
  find the use of Schoenfeld's formula to be quite reliable. This is of
  relevance for complete data on~$[0, 28]$ with $\tau=28$ as before and
  different probabilities of death~$F_2(28)$ between groups. Here, the
  approach to handle deaths before time~$28$ as censorings at day $28$ would imply
  identical (no) censoring on~$[0, 28)$, but different censoring at
  time~$28$.

  Next, analysis and sample size planning should not be guided by the required
  number of patients but by the interesting parameter. To this end, we
  reiterate that subdistributon times and, in particular, subdistributon
  hazards underly the analyses of recently COVID-19 trials as outlined
  earlier, and the Table illustrates consequences of this choice. 
  In the Table, the entries~$F_{1T}(28)$, $F_{1C}(28)$, $\theta_{SD}(28)$ and
  $OR(28)$ do not change, i.e., are assumed to be the same across all
  scenarios, but the entries for~$\theta_{ES}$ and $\theta_{ES-CE}$ do change, reflecting different
  entries $F_{2T}(28)$, $F_{2C}(28)$. To this end, it is important to note
  that~$\theta_{ES}$ and $\theta_{ES-CE}$ can be modelled freely, i.e.,
  independent of each other, but, of course, the competing event probabilities
  do not share this property. In either case, the Table illustrates that
  careful planning requires assumptions on the event-specific hazard or on the
  cumulative event probability of the competing event.

In some of the recently published randomized trials on the treatment of COVID-19 (\cite{Cao2020}, \cite{Li2020}), sample size planning was performed in terms of assumed median times to clinical improvement. Both \cite{Cao2020} and \cite{Li2020} assumed for the control group a median time to clinical improvement of 20 days and a reduction of this time to 12 days in the active treatment group. For a two-sided significance level of $\alpha = 0.05$ with a power of 80\% this resulted for the trial of \cite{Cao2020} to a total sample size of 160 patients under the assumption that 75\% of the patients would reach clinical improvement and for the trial of \cite{Li2020} to a total sample size of 200 patients under the assumption that 60\% of the patients would reach clinical improvement, both up to day 28. The proportions of patients with clinical improvement by day 28 were assumed to be different in both trials although identical median times to clinical improvement had been assumed. This could be due to different assumptions regarding the expected mortality rates not mentioned explicitly. From these specifications we speculate that an exponential distribution for time to clinical improvement had been assumed leading to an event-specific hazard ratio of 1.66 and a required number of patients experiencing the event clinical improvement of 120, which leads to the above mentioned patient numbers under the assumed proportions of clinical improvement by day 28. We note that in the statistical analysis of the trials, parameters were estimated from the subdistribution time, i.e. the subdistribution hazard ratio and median times to clinical improvement based on quantity~\eqref{eq:median}, being not quite consistent with the methods used for sample size calculation.

If the aim is to increase, say, the number of recoveries and to obtain these recoveries in a shorter time, the primary analysis may target the cumulative recovery probability as a function of time. Assuming that all or almost all patients experience one of the competing outcomes on $[0, \tau]$, it will suffice to target this probability, because an increase of the recovery probability would then protect against a harmful effect on mortality. One possibility to demonstrate both an increase of the cumulative recovery probability and a shorter time to recovery is to establish a subdistribution hazard ratio larger than one. However, the interpretation of the subdistribution hazard is not straightforward, and an alternative would be a transformation model of the cumulative recovery probability using a logistic link or a comparison of the cumulative recovery probabilities \citep{Eriksson2015} using confidence bands \citep{beyersmann2013weak}.

When the cumulative recovery probability on $[0, \tau]$ is the target parameter, we see in Table~\ref{tab:n} no large difference in the calculated sample size for the subdistribution hazard ratio based on (\ref{eq:E}) and (\ref{eq:Psi}) as compared to the calculated sample size for the odds ratio of the binary endpoint based on formula (2) in \citet{Hsieh1998}. When no competing events are present, it had been shown by \citet{Annesi1989} that efficiency of an analysis with logistic regression is high as compared to an analysis with Cox regression in the situation of a low event rate. We are not aware of a similar efficiency investigation comparing the Fine and Gray model with the logistic model in the presence of competing events. Formula (\ref{eq:jbms}) indicates that the subdistribution hazard is lower than the event-specific hazard, so arguments related to low event rates could possibly translate.

\subsection{Sample size recalculation}
As we have seen in Section \ref{sec:power}, the sample size or power calculations rely on a number of assumptions. In particular in an epidemic situation, there is no or very little prior knowledge regarding relevant parameters. These include the treatment effect but also potentially a range of nuisance parameters such as event probabilities regarding events of interest such as recovery, or competing events such as death. Sample size recalculation procedures were suggested to deal with this type of uncertainty and to make trials more robust to parameter misspecifications in the planning phase (see e.g. \citet{muetze2020} for a recent overview). Generally, two broad classes of procedures are distinguished, namely sample size recalculation based on nuisance parameters and effect-based sample size recalculation. 

Designs with sample size recalculation based on nuisance parameters are also known as internal pilot study designs \citep{friede2006}. The general procedure consists of the following steps: (i) a conventional sample size calculation is carried out at the design stage; (ii) part way through the trial the nuisance parameters are estimated from the available data and the sample size recalculated accordingly; and (iii) in the final analysis the combined sample of the internal pilot study and the remaining trial are analyzed. Nuisance parameters such as event probabilities might relate to the control group or the overall study population across the treatment groups. The latter can obviously be estimated from non-comparative data during the ongoing trial and does not require any unblinding. Therefore, it is often the preferred option, in particular in trials with regulatory relevance \citep{ema2007, fda2018}. 

With a binary outcome the overall event probability can be considered the nuisance parameter, which can be estimated from the overall sample. \citet{gould1992} provides sample size recalculation formulas based on the overall event probability for the odds ratio (considered above) as effect measure but also relative risks and risk differences. The latter was also studied in more detail by \citet{Friede2004}. The specification of the treatment effect is relevant here, since it is kept fixed. With the odds ratio as effect measure and event probabilities below (above) 0.5 a lower (higher) than expected event probability results in a sample size increase, whereas with a risk difference the sample size would be decreased. Under the proportional odds model the blinded procedure for binary outcomes can be extended to ordinal outcomes. Rather than estimating the event probability from the sampled pooled across the treatment arms the distribution of the ordinal outcome is assessed in the pooled sample \citep{bolland1998}. Guidance on blinded sample size recalculation procedures in time-to-event trials is provided in \citet{Friede2019} and references therein. In designs with flexible follow-up times, the procedures would consider the recruitment, event and censoring processes. In the situation considered here, trials are likely to use a fixed follow-up design following all patients up to $\tau$, say $\tau=28$ days. From Section \ref{sec:power} follows then that the probabilities of the event of interest and of the competing event would be estimated by the Aalen-Johansen estimator at interim. These findings would then be used to update the initial sample size calculation. 

In so-called internal pilot study designs, the sample size calculation is typically at a single time point during the study. Since the blinded procedure seems to be uncritical in terms of logistics and type I error rate inflation, repeated recalculations based on blinded data could be considered. Actually the nuisance parameters could even be monitored in a blinded fashion from a certain point in time onwards. This is also known as blinded continuous monitoring \citep{friede2012}. In fact, this is typically done in event driven trials where the total number of events across both treatment arms are monitored. This principle can be transferred to other types of outcomes such as recurrent events \citep{friede2019bcm, muetze2020bcm}. 

Group sequential designs belong to the class of designs with effect based sample size adaptation. They are used in many disease areas including oncology as well as cardiovascular and cardiometabolic research. For binary outcomes or ordinal outcomes under the proportional odds model the procedures are well established \citep{jennison2000}. \citet{logan2013} described group sequential procedures in the presence of competing events. Classical group sequential designs, however, must proceed in a prespecified manner and the size of the design stages must not be based on observed treatment effects unless prespecified weights for the design stages are used \citep{cui1999}. The latter procedure is equivalent with the inverse normal combination function by \citet{lehmacher1999}. Some issues in this type of designs with time-to-event outcomes were raised \citep{bauer2004}, but are not a concern in designs with fixed follow-up time considered here as long as the patients are grouped into design stages in the analysis\citep{friede2011}. For a very recent review on adaptive designs for COVID-19 intervention trials see \citet{Stallard2020}.

\section{Discussion} \label{sec:discussion}
In the COVID-19 pandemic the fast development of safe and effective treatments is of paramount importance. Severe forms of COVID-19 require hospitalization and in some cases intensive care. In these settings, recovery, mechanical ventilation, mortality etc. are relevant outcomes. From a statistical viewpoint different approaches to their analysis might be meaningful. Here we argued that a successful treatment of COVID-19 patients (i) increases the probability of a recovery within a certain time interval, say 28 days; (ii) aims to expedite recovery within this time frame; and (iii) does not increase mortality over this time period. We made some recommendations regarding the design and analysis of COVID-19 trials with such outcomes. Since there is no previous experience with COVID-19, sample size calculations have to be informed by data from related diseases. This results in considerable uncertainty which can be mitigated by appropriate adaptive designs including blinded sample size reestimation. 

Here we considered trials evaluating treatments of patients suffering from severe forms of COVID-19. Of course, running trials in other disease areas have been affected by the pandemic. The issues and potential solutions are discussed in a recent paper by \citet{Kunz2020}. Furthermore, we did not consider vaccine or diagnostic trials. Also, we assumed that event times were recorded on a continuous scale. In practice, however, this is strictly speaking not the case as event times might be reported in terms of days from randomization. In particular with shorter follow up times, this type of discreteness could be dealt with using appropriate models. For an overview, we defer the reader to \citet{Schmid2020}.


\noindent {\bf{Conflict of Interest}}

\noindent {\it{The authors have declared no conflict of interest.}}

\section*{Appendix}
  The aim is to show that Equation~\eqref{eq:yeah}, i.e.,
  \begin{displaymath}
    \hat P(T\le t, X(T)=j) = 1 - \prod_{u \le t}\left(1-\frac{\Delta
        N_{01}(u)}{\tilde Y(u)}\right),
  \end{displaymath}
  holds. Checking the increments of any Kaplan-Meier-type estimator, we find
  that the right-hand side of the previous display equals
  \begin{equation}
    \label{eq:curly}
    \sum_{u\le t} \left\{\prod_{v < u}\left(1-\frac{\Delta
          N_{01}(v)}{\tilde Y(v)}\right)\right\} \frac{\Delta N_{01}(u)}{\tilde Y(u)}.
  \end{equation}
  Now, $\tilde Y$ is a left-continuous `at-risk'-process which includes all
  previous deaths. Introducing
  \begin{displaymath}
    N_{02}(t-)  =  \mbox{no. of~$0\to 2$ transitions on~$(0,t)$},\\
  \end{displaymath}
  we find that the product in the curly braces of Equation~\eqref{eq:curly}
  has factors of the form
  \begin{displaymath}
    \frac{Y(v) + N_{02}(v-) - \Delta N_{01}(v)}{Y(v) + N_{02}(v-)},
  \end{displaymath}
  where~$Y$ is the usual at-risk process. Assuming no censoring on~$[0,\tau)$,
  we have that for two neighbouring type~1 event times~$v_1 < v_2$, i.e.,
  $\Delta N_{01}(v_1)\neq 0 \neq \Delta N_{01}(v_2)$ and $N_{01}(v_1) =
  N_{01}(v)$ for all $v \in [v_1, v_2)$,
  \begin{displaymath}
    Y(v_1) + N_{02}(v_1-) - \Delta N_{01}(v_1) = Y(v_2) + N_{02}(v_2-).
  \end{displaymath}
  As a consequence, canceling the appropriate terms leads to
  \begin{displaymath}
    \prod_{v < u}\left(1-\frac{\Delta
          N_{01}(v)}{\tilde Y(v)}\right)= \frac{\tilde Y(u)}{n}
  \end{displaymath}
  and
  \begin{displaymath}
    1 - \prod_{u \le t}\left(1-\frac{\Delta
        N_{01}(u)}{\tilde Y(u)}\right) = \frac{N_{01}(t)}{n},
  \end{displaymath}
  i.e., the number of type~1 events on~$[0,t]$ divided by sample size. It is
  well known that the Aalen-Johansen estimator also equals~${N_{01}(t)}/{n}$
  in the absence of censoring, which completes the argument.


\begin{thebibliography}{99}
\providecommand{\natexlab}[1]{#1}
\providecommand{\url}[1]{\texttt{#1}}
\providecommand{\urlprefix}{URL }

\bibitem[{Andersen(2013)}]{andersen2013decomposition}Andersen, P.~K. (2013). Decomposition of number of life years lost according to causes of death. \emph{Statistics in medicine} \textbf{32}, 5278--5285.

\bibitem[{Andersen and Keiding(2012)}]{andersen2011interpretability}
Andersen, P. and Keiding, N. (2012). Interpretability and importance of functionals in competing risks and multistate models. \emph{Statistics in Medicine} \textbf{31}, 1074--1088.

\bibitem[Annesi et al.(1989)]{Annesi1989}Annesi, I., Moreau, T., and Lellouch, J. (1989). Efficiency of the logistic regression and Cox proportional hazards models in longitudinal studies. \textit{Statistics in Medicine} \textbf{8}, 1515--1521.

\bibitem[Baayen et al.(2019)]{Baayen2019}Baayen, C., Volteau, C., Flamant, C., and Blanche, P. (2019). Sequential trials in the context of competing risks: Concepts and case study, with R and SAS code. \textit{Statistics in Medicine} \textbf{38}, 3682--3702.

\bibitem[Bauer and Posch(2004)]{bauer2004}Bauer, P., and Posch, M. (2004). Letter to the editor: Modification of the sample size and the schedule of interim analyses
in survival trials based on data inspections. \textit{Statistics in Medicine} \textbf{23}, 1333--1335.

\bibitem[Beigel et al.(2020)]{Beigel2020}Beigel,~J.H., Tomashek, K.M., Dodd, L.E., Mehta, A.K., Zingman, B.S., Kalil, A.C., Hohmann, E., Chu, H.Y., Luetkemeyer, A., Kline, S., Lopez de Castilla, D., Finberg, R.W., Dierberg, K., Tapson, V., Hsieh, L., Patterson, T.F., Paredes, R., Sweeney, D.A., Short, W.R., Touloumi, G., Lye, D.C., Ohmagari, N., Oh, M., Ruiz-Palacios, G.M., Benfield, T., Fätkenheuer, G., Kortepeter, M.G., Atmar, R.L., Creech, C.B., Lundgren, J., Babiker, A.G., and Pett, S.
J.D. Neaton, T.H. Burgess, T. Bonnett, M. Green, M. Makowski, A. Osinusi, S. Nayak, and H.C. Lane, for the ACTT-1 Study Group Members (2020). Remdesivir for the Treatment of Covid-19 -- Final Report. \textit{New England Journal of Medicine} \textbf{383}, 1813--1826.
	
\bibitem[Benkeser et al.(2020)]{Benkeser2020}Benkeser, D., Diaz, I., Luedtke, A., Segal, J., Scharfstein, D., and Rosenblum, M. (2020). Improving precision and power in randomized trials for COVID-19 treatments using covariate adjustment, for binary, ordinal, and time-to-event outcomes. \textit{Biometrics} (in press) doi:10.1111/biom.13377.

\bibitem[{Beyersmann et~al.(2012)Beyersmann, Allignol, and Schumacher}]{bam}Beyersmann, J., Allignol, A., and Schumacher, M. (2012). \emph{Competing Risks and Multistate Models with R}. Springer, New York.

\bibitem[{Beyersmann et~al.(2007)Beyersmann, Dettenkofer, Bertz, and Schumacher}]{j06b:_iscb07}Beyersmann, J., Dettenkofer, M., Bertz, H., and Schumacher, M. (2007). A competing risks analysis of bloodstream infection after stem-cell transplantation using subdistribution hazards and cause-specific hazards. \emph{Statistics in Medicine} \textbf{26}, 5360--5369.

\bibitem[{Beyersmann et~al.(2014)Beyersmann, Gastmeier, and Schumacher}]{beyersmann2014incidence}Beyersmann, J., Gastmeier, P., and Schumacher, M. (2014). Incidence in {ICU} populations: how to measure and report it? \emph{Intensive Care Medicine} \textbf{40}, 871--876.

\bibitem[{Beyersmann et~al.(2009)Beyersmann, Latouche, Buchholz, and Schumacher}]{beyersmann08:_simul} Beyersmann, J., Latouche, A., Buchholz, A., and Schumacher, M. (2009). Simulating competing risks data in survival analysis. \emph{Statistics in Medicine} \textbf{28}, 956--971.

\bibitem[{Beyersmann and Scheike({2014})}]{beyersmann13:_compet} Beyersmann, J. and Scheike, T. ({2014}).\emph{Handbook of Survival Analysis (Ed. {K}lein, J et al.)}, chapter Competing risks regression models. Boca Raton, FL: Chapman \& Hall/ CRC.

\bibitem[{Beyersmann and Schumacher(2008)}]{Beye:Schu:note} Beyersmann, J. and Schumacher, M. (2008). A note on nonparametric quantile inference for competing risks and
  more complex multistate models. \emph{Biometrika} \textbf{95}, 1006--1008.

\bibitem[{Bluhmki et~al.(2020)Bluhmki, Schmoor, Finke, Schumacher, Soci{\'e} \emph{et~al.}}]{bluhmki2020relapse}Bluhmki, T., Schmoor, C., Finke, J., Schumacher, M., Soci{\'e}, G., \emph{et~al.} (2020). Relapse-and Immunosuppression-Free Survival after Hematopoietic Stem Cell Transplantation: How can we assess treatment success for complex time-to-event endpoints? \emph{Biology of Blood and Marrow Transplantation} \textbf{26}, 992--997.

\bibitem[Beyersmann et al.(2013)]{beyersmann2013weak}Beyersmann, J., Termini, S.D., and Pauly, M. (2013). Weak convergence of the wild bootstrap for the Aalen--Johansen estimator of the cumulative incidence function of a competing risk. \textit{Scandinavian Journal of Statistics} \textbf{40}, 387--402.

\bibitem[Bolland et al.(1998)]{bolland1998}Bolland, K., Sooriyarachchi, M. R. and Whitehead, J. (1998). Sample size review in a head injury trial with ordered categorical responses. \textit{Statistics in Medicine} \textbf{17}, 2835--2847.

\bibitem[Cao et al.(2020)]{Cao2020}Cao, B., Wang, Y., Wen, D., Liu, W. (2020). A trial of Lopinavir–Ritonavir in adults hospitalized with severe Covid-19. \textit{New England Journal of Medicine} \textbf{382}, 1787--1799.

\bibitem[Cui et al.(1999)]{cui1999}Cui, L., Hung, H.M., and Wang, S.J. (1999). Modification of sample size in group sequential clinical trials. \textit{Biometrics} \textbf{55}: 853--857.

\bibitem[{Dodd et~al.(2020)Dodd, Follmann, Wang, Koenig, Korn \emph{et~al.}}]{dodd2020endpoints}Dodd, L.~E., Follmann, D., Wang, J., Koenig, F., Korn, L.~L., \emph{et~al.}(2020). Endpoints for randomized controlled clinical trials for COVID-19 treatments. \emph{Clinical Trials} \textbf{17}, 472--482.

\bibitem[EMA(2007)]{ema2007}EMA (2007). Reflection paper on methodological issues in confirmatory clinical trials planned with an adaptive design. London: EMEA.

\bibitem[Eriksson et al.(2015)]{Eriksson2015}Eriksson, F., Li, J., Scheike, T., and Zhang, M.J. (2015). The proportional odds cumulative incidence model for competing risks. \textit{Biometrics} \textbf{71}, 687--695.

\bibitem[FDA(2018)]{fda2018}FDA (2018). Guidance for industry: Adaptive design clinical trials for drugs and biologics. Washington DC: Food and Drug Administration.
		
\bibitem[Fine and Gray(1999)]{FineGray1999}Fine JP, Gray RJ (1999). A Proportional Hazards Model for the Subdistribution of a Competing Risk. \textit{Journal of the American Statistical Association} \textbf{94}, 496--509.
  
\bibitem[Friede and Kieser(2004)]{Friede2004}Friede,~T. and Kieser,~M. (2004). Sample size recalculation for binary data in internal pilot study designs. \textit{Pharmaceutical Statistics} \textbf{3}, 269--279.

\bibitem[Friede and Kieser(2006)]{friede2006}Friede,~T. and Kieser,~M. (2006). Sample Size Recalculation in Internal Pilot Study Designs:
A Review. \textit{Biometrical Journal} \textbf{48}, 1--19.

\bibitem[Friede et al.(2011)]{friede2011}Friede, T., Parsons, N., Stallard, N., Todd, S., Valdes Marquez, E., Chataway, J., and Nicholas, R. (2011). Designing a seamless phase II/III clinical trial using early outcomes for treatment selection: An application in multiple sclerosis. \textit{Statistics in Medicine} \textbf{30}, 1528--1540.

\bibitem[Friede et al.(2019)]{Friede2019}Friede,~T., Pohlmann,~H., and Schmidli,~H. (2019). Blinded sample size reestimation in event-driven clinical trials: Methods and an application in multiple sclerosis. \textit{Pharmaceutical Statistics} \textbf{18}, 351--365.

\bibitem[Friede et al.(2019)]{friede2019bcm}Friede,~T., H\"aring,~D.A., and Schmidli,~H. (2019). Blinded continuous monitoring in clinical trials with
recurrent event endpoints. \textit{Pharmaceutical Statistics} \textbf{18}, 54--64.

\bibitem[Friede and Miller(2012)]{friede2012}Friede, T., and Miller, F. (2012). Blinded continuous monitoring of nuisance parameters in clinical trials. \textit{Journal of the Royal Statistical Society Series C} \textbf{61}, 601--618.

\bibitem[{Geskus(2011)}]{gesk:2010} Geskus, R. (2011). Cause-specific cumulative incidence estimation and the {F}ine and {G}ray model under both left truncation and right censoring. \emph{Biometrics} \textbf{67}, 39--49.

\bibitem[Goldman et al.(2020)]{Goldman2020}Goldman, J.D., Lye, D.C.B., Hui, D.S., et al. (2020). Remdesivir for 5 or 10 days in patients with severe Covid-19. \textit{New England Journal of Medicine}, May 27, 2020. DOI: 10.1056/NEJMoa2015301.

\bibitem[Gould(1992)]{gould1992}Gould, A.L. (1992). Interim analyses for monitoring clinical trials that do not materially affect the type I error rate. \textit{Statistics in Medicine} \textbf{11}: 55--66.

\bibitem[Grein et al.(2020)]{Grein2020}Grein,~J. et al. (2020). Compassionate use of Remdesivir for patients with severe Covid-19. \textit{New England Journal of Medicine} DOI: 10.1056/NEJMoa2007016

\bibitem[{Hao et~al.(2020)Hao, Huang, Liu, Chen, Li \emph{et~al.}}]{HAO2020354}Hao, Y., Huang, L., Liu, X., Chen, Y., Li, J., \emph{et~al.} (2020). Influenza-attributable years of life lost in older adults in a subtropical city in China, 2012–2017: A modeling study based on a competing risks approach. \emph{International Journal of Infectious Diseases} \textbf{97}, 354--359.

\bibitem[{Happ et~al.(2019)Happ, Bathke, Brunner}]{Happ2019}Happ, M., Bathke, A.C., Brunner, E. (2019). Optimal sample size planning for the Wilcoxon-Mann-Whitney test. \emph{Statistics in Medicine} \textbf{38}, 363--375.

\bibitem[Hsieh et al.(1998)]{Hsieh1998}Hsieh FY, Bloch DA, Larsen MD (1998). A simple method of sample size calculation for the linear and logistic regression. \textit{Statistics in Medicine} \textbf{17}, 1623--1634.

\bibitem[Jennison and Turnbull(2000)]{jennison2000}Jennison, C., and Turnbull, B.W. (2000). \textit{Group Sequential Designs with Applications to Clinical Trials}. Boca Raton, Chapman \& Hall/CRC.

\bibitem[Kahan et al.(2020)]{Kahan2020}Kahan,~B.C., Morris,~T.P., White,~I.R., Tweed,~C.D., Cro,~S., Dahly,~D., Pham,~T.M., Esmail,~H., Babiker,~A., and Carpenter,~J.R. (2020). Treatment estimands in clinical trials of patients hospitalised for COVID-19: ensuring trials ask the right questions. Preprint available from \url{https://osf.io/7wxk9/}.
	
\bibitem[Kieser et al.(2013)]{Kieser2013}Kieser, M., Friede, T., and Gondan, M. (2013). Assessment of statistical significance and clinical relevance. \textit{Statistics in Medicine} \textbf{32}, 1707--1719.

\bibitem[Kunz et al.(2020)]{Kunz2020}Kunz, C.U., J\"orgens, S., Bretz, F., Stallard, N., Van Lancker, K., Xi, D., Zohar, S., Gerlinger, C., and Friede, T. (2020). Clinical trials impacted by the COVID-19 pandemic: Adaptive designs to the rescue? \textit{Statistics in Biopharmaceutical Research} \textbf{12}, 461--477.

\bibitem[Latouche et al.(2004)]{Latouche2004}Latouche,~A., Porcher,~R., and Chevret,~S. (2004). Sample size formula for proportional hazards modelling of competing risks. \textit{Statistics in Medicine} \textbf{23}, 3263--3274.

\bibitem[Latouche et al.(2013)]{Latouche2013}Latouche, A., Allignol, A., Beyersmann, J., et al. (2013). A competing risks analysis should report results on all cause-specific hazards and cumulative incidence functions. \textit{Journal of Clinical Epidemiology} \textbf{66}, 648-653.

\bibitem[Lee et al.(2020)]{Lee2020}Lee et al. (2020). COVID-19 mortality in patients with cancer on chemotherapy or other anticancer treatments: a prospective cohort study. \textit{Lancet} \textbf{395}: 1919--1926.

\bibitem[Lehmacher and Wassmer(1999)]{lehmacher1999} Lehmacher, W., and Wassmer, G. (1999). Adaptive sample size calculations in group sequential trials. \textit{Biometrics} \textbf{55}, 1286--1290.

\bibitem[Li et al.(2020)]{Li2020}Li et al (2020). Effect of convalescent plasma therapy on time to clinical improvement in patients with severe and life-threatening COVID-19: A randomized clinical trial. \textit{JAMA} doi:10.1001/jama.2020.10044. Published online June 3, 2020.

\bibitem[Logan and Zhang(2013)]{logan2013}Logan, B.R., and Zhang, M.J. (2013). The use of group sequential designs with common competing risks tests. \textit{Statistics in Medicine} \textbf{32}, 899--913.

\bibitem[{McCaw et~al.(2020{\natexlab{a}})McCaw, Tian, Sheth, Hsu, Kimberly \emph{et~al.}}]{mccaw2020selecting}McCaw, Z.~R., Tian, L., Sheth, K.~N., Hsu, W.-T., Kimberly, W.~T., \emph{et~al.} (2020{\natexlab{a}}). Selecting appropriate endpoints for assessing treatment effects in comparative clinical studies for COVID-19. \emph{Contemporary Clinical Trials} \textbf{97}.

\bibitem[{McCaw et~al.(2020{\natexlab{b}})McCaw, Tian, Vassy, Ritchie, Lee \emph{et~al.}}]{mccaw2020quantify}McCaw, Z.~R., Tian, L., Vassy, J.~L., Ritchie, C.~S., Lee, C.-C., \emph{et~al.}(2020{\natexlab{b}}). How to quantify and interpret treatment effects in comparative clinical studies of COVID-19. \emph{Annals of Internal Medicine}.

\bibitem[{Meller et~al.(2019)Meller, Beyersmann, and Rufibach}]{meller2019}Meller, M., Beyersmann, J., and Rufibach, K. (2019). Joint modeling of progression-free and overall survival and computation of correlation measures. \emph{Statistics in Medicine} \textbf{38}, 4270--4289.

\bibitem[{Ohneberg and Schumacher(2014)}]{ohneberg2013sample}Ohneberg, K. and Schumacher, M. (2014). \emph{Handbook of Survival Analysis (Ed. {K}lein, J et al.)}, chapter Sample Size Calculations for Clinical Trials. Boca Raton, FL: Chapman \& Hall/ CRC.

\bibitem[M\"utze and Friede(2020)]{muetze2020}M\"utze, T., and Friede, T. (2020). Sample size re-estimation. In \textit{Handbook of Statistical Methods for Randomized Controlled Trials} by Kim, Bretz, Hampson (eds).

\bibitem[M\"utze et al.(2020)]{muetze2020bcm} M\"utze, T., Salem, S., Benda, N., Schmidli, H., and Friede, T. (2020). Blinded continuous information monitoring of recurrent event endpoints with time trends in clinical trials. \textit{Statistics in Medicine} \textbf{39}, 3968--3985.

\bibitem[Noether(1987)]{Noether1987}Noether GE (1987). Sample size determination for some common nonparametric tests. \textit{Journal of the American Statistical Association} \textbf{82}, 645--647.

\bibitem[Pintilie(2002)]{Pintilie2002}Pintilie M (2002). Dealing with competing risks: testing covariates and calculating sample size. \textit{Statistics in Medicine} \textbf{21}, 3317–3324.

\bibitem[{Saadati et~al.(2018)Saadati, Beyersmann, Kopp-Schneider, and Benner}]{saadati2017prediction}Saadati, M., Beyersmann, J., Kopp-Schneider, A., and Benner, A. (2018). Prediction accuracy and variable selection for penalized cause-specific hazards models. \emph{Biometrical Journal} \textbf{60}: 288--306.

\bibitem[{Schmidt et~al.(2020)Schmidt, Hajage, Lebreton, Monsel, Voiriot\emph{et~al.}}]{schmidt2020extracorporeal}Schmidt, M., Hajage, D., Lebreton, G., Monsel, A., Voiriot, G., \emph{et~al.}(2020). Extracorporeal membrane oxygenation for severe acute respiratory distress syndrome associated with COVID-19: a retrospective cohort study. \emph{The Lancet Respiratory Medicine} .  

\bibitem[{Schoenfeld(1981)}]{schoenfeld1981asymptotic}Schoenfeld, D. (1981). The asymptotic properties of nonparametric tests for comparing survival distributions.
\emph{Biometrika} \textbf{68}, 316--319.

\bibitem[Schoenfeld(1983)]{Schoenfeld1983}Schoenfeld, D.A. (1983). Sample-size formula for the proportional-hazards regression model. \textit{Biometrics} \textbf{39}, 499--503.

\bibitem[Schoenfeld et al.(2002)]{Schoenfeld2002}Schoenfeld,~D.A. and Bernard,~G.R. for the ARDS Network (2002). Statistical evaluation of ventilator-free days as an efficacy measure in clinical trials of treatments for acute respiratory distress syndrome. \textit{Critical Care Medicine} \textbf{30}, 1772--1777.

\bibitem[Schoenfeld(2006)]{Schoenfeld2006}Schoenfeld,~D. (2006). Survival methods, including those using competing risk analysis, are not appropriate for intensive care unit outcome studies. \textit{Critical Care} \textbf{10}, 103.

\bibitem[Schmid and Berger(2020)]{Schmid2020}Schmid,~M., and Berger,~M. (2020). Competing risks analysis for discrete time-to-event data. \textit{WIREs Computational Statistics} e1529.

\bibitem[Schmoor et al.(2013)]{Schmoor2013}Schmoor,~C., Schumacher,~M., Finke,~J., and Beyersmann,~J. (2013). Competing risks and multistate models. \textit{Clinical Cancer Research} \textbf{19}, 12--21.
		
\bibitem[Schulgen et al(2005)]{Schulgen2005}Schulgen et al (2005). Sample sizes for clinical trials with time-to-event endpoints and competing risks. \textit{Contemporary Clinical Trials} \textbf{26}, 386--396.

\bibitem[Sommer et al.(2018)]{Sommer2018}Sommer,~H., Bluhmki,~T., Beyersmann,~J., and Schumacher,~M. on behalf of the COMBACTE-NET and COMBACTE-MAGNET consortium (2018). Assessing noninferiority in treatment trials for severe infectious diseases: an extension to the entire follow-up period using a cure-death multistate model. \textit{Antimicrobial Agents and Chemotherapy} \textbf{62}, e01691--17.

\bibitem[{Spinner et~al.(2020)Spinner, Gottlieb, Criner, L{\'o}pez, Cattelan \emph{et~al.}}]{spinner2020effect}Spinner, C.~D., Gottlieb, R.~L., Criner, G.~J., L{\'o}pez, J. R.~A., Cattelan, A.~M., \emph{et~al.} (2020). Effect of remdesivir vs standard care on clinical status at 11 days in patients with moderate COVID-19: a randomized clinical trial. \emph{JAMA} \textbf{324}, 1048--1057.
		
\bibitem[Stallard et al.(2020)]{Stallard2020} Stallard, N., Hampson, L., Benda, N., Brannath, W., Burnett, T., Friede, T., Kimanim P.K., Koenig, F., Krisam, J., Mozgunov P., Posch, M., Wason, J., Wassmer, G., Whitehead, J., Williamson, S.F., Zohar, S., and Jaki, T. (2020). Efficient adaptive designs for clinical trials of interventions for COVID-19. \textit{Statistics in Biopharmaceutical Research} \textbf{12}, 483--497.
		
\bibitem[Tai et al.(2018)]{Tai2015}Tai et al (2018). Estimating sample size in the presence of competing risks – Cause-specific hazard or cumulative incidence approach? \textit{Statistical Methods in Medical Research} \textbf{27}, 114--125.

\bibitem[{Tan(2020)Tan}]{Tan2020}Tan, K.S. (2020). Letter: A concern about survival time as an endpoint in coronavirus disease 2019 clinical trials. \emph{Clinical Trials} (in press).

\bibitem[{Vollandt and Horn(1997)Vollandt and Horn}]{Vollandt1997} Vollandt, R., and Horn, M. (1997). Evaluation of Noether's method of sample size determination for the Wilcoxon-Mann-Whitney test. \emph{Biometrical Journal} \textbf{39}, 823--829.

\bibitem[Wang et al.(2020)]{Wang2020}Wang, Y. et al. (2020). Remdesivir in adults with severe COVID-19: a randomised, double-blind, placebo-controlled,
  multicentre trial. \textit{Lancet} DOI:10.1016/S0140-6736(20)31022-9
	
\bibitem[Whitehead(1993)]{Whitehead1993}Whitehead, J. (1993). Sample size calculations for ordered categorical data. \textit{Statistics in Medicine} 12, 2257-2271.

\bibitem[Wilt et al.(2020)]{Wilt2020}Wilt, T.J., Kaka, A.S., MacDonald, R., Greer, N., Obley, A., and Duan-Porter, W. (2020). Remdesivir for adults with COVID-19:
A living systematic review for an American College of Physicians Practice Points. \textit{Annals of Internal Medicine} \url{doi:10.7326/M20-5752}.

\bibitem[WHO(2020)World Health Organization]{who2020}World Health Organization (2020). COVID-19 Therapeutic Trial Synopsis. \url{https://www.who.int/publications/i/item/covid-19-therapeutic-trial-synopsis} (accessed 22 NOV 2020).

\bibitem[Yehya et al.(2019)]{Yehya2019}Yehya, N. et al. (2019). Reappraisal of ventilator-free days in critical care research. \textit{American Journal of Respiratory and Critical Care Medicine} \textbf{2000}, 828--836.

\end{thebibliography}
\end{document}